\documentclass[12pt]{iopart}
\usepackage{amssymb}
\usepackage{graphicx}
\usepackage{color}

\def\beq{\begin{eqnarray}}
\def\eeq{\end{eqnarray}}

\begin{document}

\title{Variational collocation for systems of coupled anharmonic oscillators}
\author{Paolo Amore}
\address{Facultad de Ciencias, CUICBAS, Universidad de Colima, \\
Bernal D\'{\i}az del Castillo 340, Colima, Colima, M\'exico} 

\author{Francisco M. Fern\'andez}
\address{INIFTA (UNLP, CCT La Plata-CONICET), Divisi\'{o}n Qu\'{i}mica
Te\'{o}rica, Diag. 113 y 64 (S/N), Sucursal 4, Casilla de Correo 16, 1900 La
Plata, Argentina}

\begin{abstract}
We have applied a collocation approach to obtain the numerical solution to
the stationary Schr\"odinger equation for systems of coupled oscillators.
The dependence of the discretized Hamiltonian on scale and angle parameters
is exploited to obtain optimal convergence to the exact results. A careful
comparison with results taken from the literature is performed, showing the
advantages of the present approach.
\end{abstract}

\maketitle



\section{INTRODUCTION}

\label{sec:intro}

Coupled anharmonic oscillators have proved useful for modelling a wide
variety of physical problems, such as, for example, the vibrations of
polyatomic molecules\cite{C76,CSK78,CH86}. For this reason there has been
great interest in the calculation of their eigenvalues and eigenfunctions%
\cite
{Hioe78,FO93,RDF97,RDF98,RDF99,DF01,PE81,PE81b,Witwit93,Kaluza,BSP96,FMPZM99,CC07}
(and references therein).

In this paper we propose an alternative variational method for the
calculation of eigenvalues and eigenfunctions of coupled anharmonic
oscillators. We develop the approach in Sec.~\ref{sec:method}, in Sec.~\ref
{sec:HO} we consider four coupled harmonic oscillators that are useful for
testing our approach because the model is exactly solvable. In sections~\ref
{sec:D=2}, \ref{sec:D=3} and \ref{sec:D=4} we obtain the eigenvalues of two,
three and four coupled anharmonic oscillators, respectively. In each case we
compare our results with those obtained earlier by other authors using
different approaches. Finally, in Sec.~\ref{sec:conclusions} we draw
conclusions.

\section{THE METHOD}

\label{sec:method}

In order to obtain accurate eigenvalues and eigenfunctions of the
Schr\"{o}dinger equation for coupled anharmonic oscillators we propose a
collocation approach that allows the discretization of a $D$--dimensional
region by means of a particular set of functions called \textsl{Little Sinc
functions} (LSF)~\cite{Amore07a,Amore07b,Amore08,Amore09a,Amore09b}. They
were proposed by Baye\cite{Baye95} who called them ``first sine basis''.
This basis set proved useful for the accurate variational treatment of the
one--dimensional Schr\"{o}dinger equation\cite{Amore07a}, the representation
of non--local operators on a uniform grid for the solution of the
relativistic Salpeter equation \cite{Amore07b}, and for the accurate
treatment of the Helmholtz equation on arbitrary two--dimensional domains,
both for the homogeneous \cite{Amore08} and inhomogeneous \cite{Amore09a}
case. Recently, we investigated the practical utility of four sets of LSF%
\cite{Amore09b} with different boundary conditions that include the original
set\cite{Amore07a} as a particular case. Here, we restrict ourselves to the
LSF set with Dirichlet boundary conditions\cite{Amore07a}.

To make our discussion self--contained we outline the main features of our
approach, starting with the set of functions which are used for
discretization. A LSF is an approximate representation of the Dirac delta
function in terms of the wave functions of a particle in a box of size $2L$:
\begin{eqnarray}
s_{k}(h,N,x) &\equiv &\frac{1}{2N}\ \left\{ \frac{\sin \left( (2N+1)\ \chi
_{-}(x)\right) }{\sin \chi _{-}(x)}\right.  \nonumber \\
&-&\left. \frac{\cos \left( (2N+1)\chi _{+}(x)\right) }{\cos \chi _{+}(x)}%
\right\} \ ,  \label{sincls}
\end{eqnarray}
where $\chi _{\pm }(x)\equiv \frac{\pi }{2Nh}(x\pm kh)$. The index $k$ takes
all the integer values between $-N/2+1$ and $N/2-1$, where $N$ is an even
integer. The LSF $s_{k}$ is peaked at $x_{k}=2Lk/N=kh$, $h$ being the grid
spacing and $2L$ the total extension of the interval where the function is
defined. These LSF satisfy $s_{k}(h,N,x_{j})=\delta _{kj}$ and are
orthogonal
\[
\int_{-L}^{L}s_{k}(h,N,x)s_{j}(h,N,x)dx=h\ \delta _{kj}\ .
\]

It follows from those properties of the LSF that we can approximate a
function defined on $x\in (-L,L)$ as
\begin{equation}
f(x)\approx \sum_{k=-N/2+1}^{N/2-1}f(x_{k})\ s_{k}(h,N,x)\ . \label{f(x)_LSF}
\end{equation}
In a similar way we can also obtain a representation of the derivatives of a
LSF as:
\begin{eqnarray}
\frac{ds_{k}(h,N,x)}{dx} &\approx &\sum_{j}\left. \frac{ds_{k}(h,N,x)}{dx}%
\right| _{x=x_{j}}\ s_{j}(h,N,x)  \nonumber \\
&\equiv &\sum_{j}c_{kj}^{(1)}\ s_{j}(h,N,x)  \label{der1} \\
\frac{d^{2}s_{k}(h,N,x)}{dx^{2}} &\approx &\sum_{j}\left. \frac{%
d^{2}s_{k}(h,N,x)}{dx^{2}}\right| _{x=x_{j}}\ s_{j}(h,N,x)  \nonumber \\
&\equiv &\sum_{j}c_{kj}^{(2)}\ s_{j}(h,N,x)\ ,  \label{der2}
\end{eqnarray}
where the analytical expressions for the coefficients $c_{kj}^{(r)}$ have
been given elsewhere\cite{Amore07a}.

Although Eq.~(\ref{f(x)_LSF}) is not exact, we can make the error of that
representation of the function $f(x)$ as small as possible by simply
increasing the value of $N$, as discussed in our earlier paper\cite{Amore07a}%
. The effect of this approximation is the discretization of the continuous
interval $2L$ into a set of $N-1$ uniformly spaced points, $x_{k}$. For
example, the application of this approach to a one--dimensional eigenvalue
problem results in the diagonalization of a $(N-1)\times (N-1)$ matrix.

An appropriate basis set for a $D$--dimensional problem is given by the
direct product of one--dimensional LSF that generates a uniform grid with
spacing $h$ (in some particular cases it may be more convenient to consider
different spacing in different directions). A set of $D$ integers $%
(k_{1},k_{2},\dots ,k_{D})$ completely specifies the location of a given
point inside the hyper--volume. However, with the purpose of constructing
Hamiltonian matrices it is convenient to identify one such point with just a
single integer $\bar{K}_{D}$ that takes all the values between $1$ and $%
(N-1)^{D}$ as shown in the Appendix~\ref{sec:matrix}

In this paper we only consider Hamiltonian operators of the form
\begin{equation}
\hat{H}=T(\hat{p}_{1},\hat{p}_{2},\ldots ,\hat{p}_{D})+V(\hat{x}_{1},\hat{x}%
_{2},\ldots ,\hat{x}_{D})
\end{equation}
where $T$ is a polynomial of second degree and $V$ is a polynomial function
of the coordinates that is bounded from below. We assume that the
eigenfunctions $\psi _{n}(x_{1},\dots ,x_{D})$ are defined in a $D$%
--dimensional hypercube $\Omega $ of side $L$ and satisfy Dirichlet boundary
conditons on the frontier $\partial \Omega $.

By means of the discretization based on the direct product of LSF outlined
above we obtain a Hamiltonian matrix of the form
\begin{eqnarray}
H_{k_{1}\dots k_{D},k_{1}^{\prime }\dots k_{D}^{\prime }}
&=&T_{k_{1}k_{1}^{\prime }}\delta _{k_{2}k_{2}^{\prime }}\dots \delta
_{k_{D}k_{D}^{\prime }}\\ \nonumber
&+&\ldots +\delta _{k_{1}k_{1}^{\prime }}\dots \delta
_{k_{D-1}k_{D-1}^{\prime }}T_{k_{D}k_{D}^{\prime }}  \nonumber \\
&+&V(x_{k_{1}},\dots ,x_{k_{D}})\ \delta _{k_{1}k_{1}^{\prime }}\dots \delta
_{k_{D}k_{D}^{\prime }}  \label{Hamiltonian}
\end{eqnarray}
where $k_{i}$ and $k_{i}^{\prime }$ range from $-N/2+1$ to $N/2-1$. Arguing
as in Appendix~\ref{sec:matrix} we express the $2D$ labels $k_{i}$
and $k_{i}^{\prime }$ in terms
of just two $\bar{K}_{D}$ and $\bar{K}_{D}^{\prime }$ and obtain a $%
(N-1)^{D}\times (N-1)^{D}$ matrix representation of the Hamiltonian
operator. We expect that its eigenvalues and eigenvectors tend to the
energies and wave functions, respectively, of the Hamiltonian operator as $N$
increases. Notice that the potential part of this matrix is \textsl{diagonal}%
, while the kinetic one is sparse. These two features facilitate the
numerical treatment of multidimensional problems as we will see in what follows.

In principle we have two adjustable parameters: $N$ and $L$, but we can bind
them together by means of the variational method. As discussed elsewhere\cite
{Amore07a} it is convenient to set the optimal value of the scale parameter $%
L$ in such a way that the trace of the Hamiltonian matrix is minimum. Since
the Hamiltonian matrix is a relatively simple analytical function of $L$ the
calculation just indicated does not offer any dificulty and we obtain the
optimal scale parameter as an analytic function of $N$: $L_{PMS}(N)$. Here,
PMS stand for \textsl{principle of minimal sensitivity}\cite{Ste81} and the resulting
Hamiltonian matrix depends only on $N$. The construction of the
kinetic--energy matrix $\mathbf{T}$ is the time--consuming part of the
process of building the Hamiltonian matrix $\mathbf{H}$. However, in the
problems discussed here the form of $\mathbf{T}$ depends only on $N$ and $D$
and is suitable for several models with different potential--energy
functions $V$. We can thus take advantage of the fact that the calculation
of the matrix $\mathbf{V}$ is faster because it is diagonal with only $%
(N-1)^{D}$ elements.

\section{COUPLED HARMONIC OSCILLATORS \label{sec:HO}}

In order to test the accuracy and rate of convergence of our approach we
first consider a set of $D$ coupled harmonic oscillators given by the
Hamiltonian operator
\begin{equation}
H=\sum_{i}\left( -\frac{1}{2}\frac{\partial ^{2}}{\partial x_{i}^{2}}+\frac{1%
}{2}x_{i}^{2}\right) +\frac{1}{2}\sum_{i}\sum_{j}v_{ij}x_{i}x_{j}
\end{equation}
where $v_{ij}=v_{ji}$. One can easily solve the Schr\"{o}dinger equation and
obtain the eigenvalues exactly:
\begin{equation}
E_{\mathbf{n}}=\sum_{i}\sqrt{1+\lambda _{i}}\left( n_{i}+\frac{1}{2}\right)
\label{exactsho}
\end{equation}
where $n_{i}=0,1,\ldots $, $i=1,2,\ldots ,D$ are the harmonic oscillator
quantum numbers and $\lambda _{i}$ are the eigenvalues of the symmetrical
matrix with elements $v_{ij}$.

The calculation is simple, we first construct the Hamiltonian matrix and
choose the value of $L$ that makes its trace a minimum\cite{Amore07a}. Then
we obtain the eigenvalues of the resulting matrix for increasing values of $%
N $. Table~\ref{tab0} shows the energies of the ground and first excited
states of $D=4$ harmonic oscillators with $v_{ij}=(1-\delta _{ij})/3$ for
increasing matrix dimension. We appreciate that the rate of convergence of
present method is satisfactory for the treatment of coupled oscillators.
With that set of potential parameters we obtain $\lambda _{1}=1$ and $%
\lambda _{2}=\lambda _{3}=\lambda _{4}=-1/3$ so that the first excited state
shown in Table~\ref{tab0} is three--fold degenerate.

\begin{table}[tbp]
\caption{First two energy levels of the coupled harmonic oscillators.}
\label{tab0}%
\begin{tabular}{ccc}
$N^4 \times N^4$ & $E_{0}$ & $E_{1}$  \\
\hline
$5^4 \times 5^4$   & 1.929802495 & 2.755212192 \\
$7^4 \times 7^4$   & 1.931801216 & 2.748861410 \\
$9^4 \times 9^4$   & 1.931851103 & 2.748382295 \\
$11^4 \times 11^4$ & 1.931851707 & 2.748350435 \\
$13^4 \times 13^4$ & 1.931851659 & 2.748348376 \\
$15^4 \times 15^4$ & 1.931851653 & 2.748348243 \\
\hline
exact              & 1.931851653 & 2.748348234 \\
\end{tabular}
\bigskip\bigskip
\end{table}

\begin{table}[tbp]
\caption{Energies of three uncoupled harmonic oscillators}
\label{tab0b}%
\begin{tabular}{cc}
Dimension & $E_{0}$  \\
\hline
$9^3 $   &  1.4999927163457656  \\
$15^3$   & 1.4999999993138868   \\
$19^3$   & 1.4999999999986033   \\
$29^3$ &   1.5000000000000016   \\
\end{tabular}
\bigskip\bigskip
\end{table}

Some time ago, Tymczak and Wang\cite{TW97} calculated the eigenvalues of a
three--dimensional harmonic oscillator ($D=3$, $v_{ij}=0$) by means of
different basis sets of wavelets. Table~\ref{tab0b} shows that our results
for the same model are considerably more accurate than theirs.

\section{TWO COUPLED ANHARMONIC OSCILLATORS \label{sec:D=2}}

In this section we apply our approach to two coupled anharmonic oscillators
considered earlier by other authors.

\subsection{PULLEN--EDMONDS HAMILTONIAN}

Our first example of two--dimensional anharmonic oscillator is the
so--called Pullen--Edmonds Hamiltonian~\cite{PE81, PE81b}
\begin{equation}
\hat{H}=-\frac{1}{2}\nabla ^{2}+\frac{1}{2}\left( x^{2}+y^{2}\right) +\kappa
x^{2}y^{2}\ .  \label{eq:P-E-H}
\end{equation}
Proceding as indicated above we obtain the eigenvalues of the Hamiltonian
matrix, and Table \ref{tabPE2} shows the first four of them for $\kappa =1$.
We clearly appreciate that the variational eigenvalues converge reasonably
fast as $N$ increases.

\begin{table}[tbp]
\caption{Variational collocation approach to the Pullen-Edmonds hamiltonian
for $\kappa=1$. }
\label{tabPE2}%
\begin{tabular}{ccccc}
$N$ & $E_0$  & $E_1$  & $E_2$  & $E_3$  \\
\hline
20 & 1.169791833 & 2.438995552 & 2.438995552 & 3.476809761 \\
30 & 1.169783302 & 2.438859138 & 2.438859138 & 3.475378334 \\
40 & 1.169783112 & 2.438854966 & 2.438854966 & 3.475320052 \\
50 & 1.169783105 & 2.438854795 & 2.438854795 & 3.475317137 \\
60 & 1.169783105 & 2.438854786 & 2.438854786 & 3.475316964 \\
70 & 1.169783105 & 2.438854785 & 2.438854785 & 3.475316952 \\
\end{tabular}
\bigskip \bigskip
\end{table}

Some time ago by Fessatidis and collaborators~\cite{FMPZM99} chose the
Pullens--Edmons Hamiltonian~(\ref{eq:P-E-H}) with $\kappa =1$ to test their
proposed variational approach. The results obtained by those authors for
the ground state, reported in their Table I, clearly converge to a limit
that is greater than the one obtained here by means of the LSF. The reason
of the erroneous results of Fessatidis et al \cite{FMPZM99} lies not in the
method used to solve the problem, but in the fact that the authors have
resorted to an unsuitable spherically--symmetric basis set $\{\varphi _{j}(r)\}$%
, where $r=\sqrt{x^{2}+y^{2}}$. Since the eigenfunctions of $\hat{H}$ depend
on two variables, for example $\psi _{n}(x,y)$ or $\psi _{n}(r,\phi )$, $%
x=r\cos \phi $, $y=r\sin \phi $, then the spherical--symmetric basis
generated by Fessatidis et al\cite{FMPZM99} is not complete and their
eigenvalues do not converge to those of the Hamiltonian operator (\ref
{eq:P-E-H}). In fact, they obtained the eigenvalues of an effective
central--field Hamiltonian operator in which the average
\begin{equation}
\frac{1}{2\pi }\int_{0}^{2\pi }x^{2}y^{2}\,d\phi =\frac{r^{4}}{8}
\end{equation}
substitutes the anisotropic part of the potential in Eq. (\ref{eq:P-E-H}):
\begin{equation}
\hat{H}=-\frac{1}{2}\nabla ^{2}+\frac{1}{2}r^{2}+\frac{\kappa }{8}r^{4}
\label{eq:PE_rad}
\end{equation}
We have verified that the ground--state eigenvalue of this operator for $%
\kappa =1$ is $E_{0}=1.1790711996155152844$.

\subsection{QUARTIC OSCILLATORS}

Our second example of two coupled anharmonic oscillators is given by
\begin{eqnarray}
\hat{H} &=&\frac{\hat{p}_{1}^{2}}{2m_{1}}+\frac{1}{2}m_{1}\omega
_{1}^{2}x_{1}^{2}+\frac{\hat{p}_{2}^{2}}{2m_{2}}+\frac{1}{2}m_{2}\omega
_{2}^{2}x_{2}^{2}  \nonumber \\
&+&\lambda \left(
c_{40}x_{1}^{2}+c_{04}x_{2}^{4}+c_{22}x_{1}^{2}x_{2}^{2}\right) \ ,
\end{eqnarray}
that has been studied earlier by Hioe et al\cite{Hioe78} and Chung and Chew%
\cite{CC07} for $\hbar =m_{1}=m_{2}=\omega _{1}=\omega _{2}=c_{40}=c_{04}=1$
and $c_{22}=2$. Table \ref{tab1} shows present results for the first three
energy eigenvalues with different values of $\lambda $ and $N=20$. The
reader may compare our results with those contained in Tables I and II of
Ref.\cite{CC07}, which also report the results of Hioe et al \cite{Hioe78}.
It is worth noticing that we can always take into accout the symmetry of the
problem to decrease considerably the computational load. The eigenstates of
the Hamiltonian shown above have definite parity and, for example, we obtain
the results for the even-even states shown Table \ref{tab1} by means of just
$10^{2}\times 10^{2}$ matrices. Whenever possible the use of properly
symmetrized basis sets of LSF is advisable in the case of large $D$ and/or $%
N $.

\begin{table}[tbp]
\caption{Energy eigenvalues of the system of two coupled quartic anharmonic
oscillators for $\hbar =m_{1}=m_{2}=\omega _{1}=\omega _{2}=c_{40}=c_{04}=1$
and $c_{22}=2$. We use a grid with $N=20$ which yields $19^{2}\times 19^{2}$
matrices.}
\label{tab1}%
\begin{tabular}{cccc}
$\lambda$ & $E_{0}$ & $E_{1}$  & $E_{2}$ \\
\hline
$0.05$    &  1.084298606  & 2.238800191 & 3.454166066  \\
$0.1$     &  1.150188128  & 2.414340361 & 3.772322621 \\
$0.5$     &  1.476025071  & 3.231453204 & 5.195313797 \\
$1$       &  1.724184113  & 3.830324193 & 6.213815314 \\
$10$      &  3.301210724  & 7.527044432 & 12.39681625 \\
$100$     &  6.911899705  & 15.86897394 & 26.23624148 \\
$5000$    &  25.27402386  & 58.13369977 & 96.21028659 \\
\hline
$0.05$    &  1.084298606  & 2.238800180 & 3.454166056 \\
$0.1$     &  1.150188125  & 2.414340327 & 3.772322591 \\
$0.5$     &  1.476025046  & 3.231453000 & 5.195313648 \\
$1$       &  1.724184069  & 3.830323856 & 6.213815078 \\
$10$      &  3.301210571  & 7.527043378 & 12.39681556 \\
$100$     &  6.911899338  & 15.86897147 & 26.23623988 \\
$5000$    &  25.27402247  & 58.13369048 & 96.21028060 \\
\end{tabular}
\bigskip \bigskip
\end{table}

\section{THREE COUPLED ANHARMONIC OSCILLATORS \label{sec:D=3}}

\subsection{QUARTIC OSCILLATORS}

As an example of a three--dimensional anharmonic oscillator we choose the
model studied by Witwit\cite{Witwit93} some time ago:
\begin{eqnarray}
V(x,y,z) &=&\frac{1}{2}(x^{2}+y^{2}+z^{2})+\lambda \left(
a_{xx}x^{4}+a_{yy}y^{4}+a_{zz}z^{4}\right.  \nonumber \\
&+&\left. 2a_{xy}x^{2}y^{2}+2a_{yz}y^{2}z^{2}+2a_{xz}x^{2}z^{2}\right) \ .
\end{eqnarray}
His Table II shows the first eight eigenvalues for $a_{xx}=\frac{1}{2}$, $%
a_{yy}=\frac{1}{3}$, $a_{zz}=\frac{1}{6}$, $a_{xy}=a_{xz}=\frac{1}{2}$, $%
a_{yz}=\frac{1}{4}$ and several values of $\lambda $. Here we restrict
ourselves to the most unfavourable case $\lambda =10^{6}$.

We have applied our collocation method to this problem following three
approaches. The first approach is the one followed so far, which uses the
minimization of the trace of the Hamiltonian matrix to generate the optimal
scale $L$ for a given grid. The results in Table~\ref{tab2a} have been
obtained in this straightforward way that does not take
into account the anisotropy of the potential. In the second approach we take
into account that anisotropy by simply rescaling the $y$ and $z$
coordinates, $y\rightarrow \beta y$ and $z\rightarrow \gamma z$, with two
adjustable parameters $\beta $ and $\gamma $. Thus, the resulting eigenvalue
equation becomes
\begin{eqnarray}
&&-\frac{1}{2}\left[ \frac{\partial ^{2}}{\partial x^{2}}+\beta ^{2}\frac{%
\partial ^{2}}{\partial y^{2}}+\gamma ^{2}\frac{\partial ^{2}}{\partial z^{2}%
}\right] \psi (x,y/\beta ,z/\gamma )  \nonumber \\
&=&(E-V(x,y/\beta ,z/\gamma ))\psi (x,y/\beta ,z/\gamma )\ .
\end{eqnarray}
As a result, now the trace of the Hamiltonian matrix explicitly depends upon
$\beta $, $\gamma $ and $L$, which are chosen to minimize it. Notice that
this approach is equivalent to choosing LSF with different length scales on
the three axes, say $L_{x}$, $L_{y}$ and $L_{z}$. Table \ref{tab2b} shows
the convergence of the eigenvalues obtained in this way as the matrix
dimension increases.

The third approach consists of introducing additional adjustable parameters
by means of a coordinate rotation of the form
\begin{eqnarray}
x^{\prime } &=&x\cos \theta _{1}\ \cos \theta _{2}-y\sin \theta _{1}-z\ \cos
\theta _{1}\sin \theta _{2}  \nonumber \\
y^{\prime } &=&x\sin \theta _{1}\ \cos \theta _{2}+y\cos \theta _{1}-z\ \sin
\theta _{1}\sin \theta _{2}  \nonumber \\
z^{\prime } &=&x\sin \theta _{2}+z\cos \theta _{2}\ .  \label{eq:coord_rot}
\end{eqnarray}
so that the trace of the Hamiltonian matrix now depends on $L$, $\beta $,
$\gamma $, $\theta _{1}$ and $\theta _{2}$. Only the potential part of the
matrix will depend on the rotation angles because the kinetic energy is
invariant under such transformation. Table~\ref{tab2c} shows the convergence
of the eigenvalues obtained in this way as the matrix dimension increases.

Table~\ref{tab2d} shows the values of the optimal parameters for the three
methods outlined above. In Appendix~\ref{sec:Var_Met} we outline some
features of the variational method and suggest that the LSF exhibit a
variational behavior that is different from that of a basis set of
harmonic--oscillator eigenfunctions.

\begin{table}[tbp]
\caption{Energy eigenvalues of the three dimensional anharmonic oscillator $%
V(x,y,z)= \frac{1}{2} (x^2+y^2+z^2) + \lambda (a_{xx} x^4+ a_{yy} y^4+a_{zz}
z^4 + 2 a_{xy} x^2 y^2 + 2 a_{yz} y^2 z^2 + 2 a_{xz} x^2 z^2)$. $\lambda =
10^6$ and $a_{xx} = \frac{1}{2}$, $a_{yy}=\frac{1}{3}$, $a_{zz}=\frac{1}{6}$%
, $a_{xy} = a_{xz} = \frac{1}{2}$ and $a_{yz}=\frac{1}{4}$.}
\label{tab2a}%
\begin{tabular}{cccccc}
$N$ & $19^3 \times 19^3$  & $29^3 \times 29^3$  & $39^3 \times 39^3$  & Ref.\cite{Witwit93} \\
\hline
$E_0$ & 169.2157495 & 169.2145773 & 169.2145661 & 169.23\\
$E_1$ & 294.4522990 & 294.4365531 & 294.4363754 & 294.42\\
$E_2$ & 315.2612020 & 315.2602658 & 315.2602614 & 315.28\\
$E_3$ & 339.6044054 & 339.6041638 & 339.6041624 & 339.66\\
$E_4$ & 436.2738801 & 436.1607904 & 436.1591660 & -\\
$E_5$ & 456.4724743 & 456.4654890 & 456.4654254 & 456.46\\
$E_6$ & 487.7693071 & 487.7639023 & 487.7639454 & -\\
$E_7$ & 492.8611895 & 492.8571862 & 492.8570397 & 492.85\\
$E_8$ & 509.1325800 & 509.1322591 & 509.1323064 & 509.14\\
$E_9$ & 548.6572531 & 548.6517620 & 548.6516792 & -\\
\end{tabular}
\bigskip \bigskip
\end{table}

\begin{table}[tbp]
\caption{Same as Table \ref{tab2a} using the second approach.}
\label{tab2b}%
\begin{tabular}{cccccc}
$N$ & $19^3 \times 19^3$  & $29^3 \times 29^3$  & $39^3 \times 39^3$  & Ref.\cite{Witwit93} \\
\hline
$E_0$ & 169.2146979 & 169.2145663  & 169.2145660 & 169.23\\
$E_1$ & 294.4375151 & 294.4363709  & 294.4363667 & 294.42\\
$E_2$ & 315.2605725 & 315.2602620  & 315.2601985 & 315.28\\
$E_3$ & 339.6047811 & 339.6041637  & 339.6041624 & 339.66\\
$E_4$ & 436.1685952 & 436.1591847  & 436.1591447 & -\\
$E_5$ & 456.4659709 & 456.4654324  & 456.4654243 & 456.46\\
$E_6$ & 487.7665445 & 487.7638786  & 487.7638732 & -\\
$E_7$ & 492.8575850 & 492.8576386  & 492.8570724 & 492.85\\
$E_8$ & 509.1326192 & 509.1323070  & 509.1323014 & 509.14\\
$E_9$ & 548.6570096 & 548.6516923  & 548.6516780 & -\\
\end{tabular}
\bigskip \bigskip
\end{table}

\begin{table}[tbp]
\caption{Same as Table \ref{tab2a} using the third approach.}
\label{tab2c}%
\begin{tabular}{cccccc}
$N$ & $19^3 \times 19^3$  & $29^3 \times 29^3$  & $39^3 \times 39^3$
& Ref.\cite{Witwit93} \\
\hline
$E_0$ & 169.2146303 & 169.2145660  & 169.2145660  & 169.23\\
$E_1$ & 294.4368237 & 294.4363675  & 294.4363668  & 294.42\\
$E_2$ & 315.2605587 & 315.2602619  & 315.2602616  & 315.28\\
$E_3$ & 339.6043482 & 339.6041626  & 339.6041623  & 339.66\\
$E_4$ & 436.1613515 & 436.1591490  & 436.1591446  & -\\
$E_5$ & 456.4664315 & 456.4654261  & 456.4654249  & 456.46\\
$E_6$ & 487.7652490 & 487.7638755  & 487.7642896 & -\\
$E_7$ & 492.8588268 & 492.8571556  & 492.8571528 & 492.85\\
$E_8$ & 509.1332413 & 509.1323079  & 509.1323066  & 509.14\\
$E_9$ & 548.6527375 & 548.6516798  & 548.6516781  & -\\
\end{tabular}
\bigskip \bigskip
\end{table}

\begin{table}[tbp]
\caption{Optimal parameters for the problems of Tables \ref{tab2a}, \ref
{tab2b} and \ref{tab2c} .}
\label{tab2d}%
\begin{tabular}{ccccccc}
$N$ & $L_{PMS}$  & $\beta$  & $\gamma$  & $\theta_1$ & $\theta_2$ \\
\hline
$20$ & 0.2922 & - & -  & - & - \\
$30$ & 0.3309 & - & -  & - & - \\
$40$ & 0.3623 & - & -  & - & - \\
\hline
$20$ & 0.2728 & 0.91937 & 0.86287  & - & - \\
$30$ & 0.3090 & 0.91939 & 0.86283  & - & - \\
$40$ & 0.3383 & 0.91939 & 0.86281  & - & - \\
\hline
$20$ & 0.2964 & 1.01726 &  1 & 0.48115 & 0.78540 \\
$30$ & 0.3356 & 1.01727 &  1 & 0.48152 & 0.78540 \\
$40$ & 0.3674 & 1.01727 &  1 & 0.48164 & 0.78540 \\
\end{tabular}
\bigskip \bigskip
\end{table}

\subsection{SEXTIC OSCILLATOR}

Our second example of three--dimensional anharmonic oscillator is given by
the Hamiltonian operator
\begin{eqnarray}
H &=&\frac{1}{2}\left( p_{1}^{2}+p_{2}^{2}+p_{3}^{2}\right) +\frac{1}{2}%
\left( x_{1}^{2}+x_{2}^{2}+x_{3}^{2}\right)  \nonumber \\
&+&\left[ 2\left( x_{1}^{4}+x_{2}^{4}+x_{3}^{4}\right) +\frac{1}{2}\left(
x_{1}^{6}+x_{2}^{6}+x_{3}^{6}\right) \right]  \nonumber \\
&+&x_{1}x_{2}+x_{1}x_{3}+x_{2}x_{3}\ .
\end{eqnarray}
studied Braun et al\cite{BSP96} and Chung and Chew\cite{CC07}. Table \ref
{tab2} shows our numerical results for several matrix dimensions and a
variationally optimized grid scale $L$. We show the eigenvalues obtained by
Braun et al \cite{BSP96} and Chung and Chew\cite{CC07} for comparison. The
last row of Table \ref{tab2} shows the most accurate eigenvalues obtained by
Chung and Chew \cite{CC07} with a matrix of size $17^{3}\times 17^{3}$.

\begin{table}[tbp]
\caption{Energy eigenvalues of the system of three coupled sextic anharmonic
oscillators}
\label{tab2}%
\begin{tabular}{cccc}
$N$ & $E_{0}$ & $E_{1}$  & $E_{2}$ \\
\hline
$5^3 \times 5^3$    &  2.973116328 &  5.292534159  &  5.859553533  \\
$9^3 \times 9^3$    &  2.978379470 &  5.296297359  &  5.866068948  \\
$17^3 \times 17^3$  &  2.978302843 &  5.295993128  &  5.865822825  \\
$19^3 \times 19^3$  &  2.978302696 &  5.295992510  &  5.865822333  \\
$21^3 \times 21^3$  &  2.978302665 &  5.295992375  &  5.865822226  \\
$29^3 \times 29^3$  &  2.978302657 &  5.295992339  &  5.865822193  \\
Ref.\cite{BSP96}    &  2.978302    &  5.295992    &  5.865822     \\
Ref.\cite{CC07} &  2.978305    &  5.296000     &  5.865828     \\
\end{tabular}
\bigskip \bigskip
\end{table}

\section{FOUR COUPLED SECTIC OSCILLATOR \label{sec:D=4}}

As an example of four--dimensional anharmonic oscillator we choose the model
studied by Kaluza\cite{Kaluza} and Chung and Chew\cite{CC07}:
\begin{eqnarray}
H &=&\frac{1}{2}\left( p_{1}^{2}+p_{2}^{2}+p_{3}^{2}+p_{4}^{2}\right) +\frac{%
1}{2}\left( x_{1}^{2}+x_{2}^{2}+x_{3}^{2}+x_{4}^{2}\right)  \nonumber \\
&+&\left[ 2\left( x_{1}^{4}+x_{2}^{4}+x_{3}^{4}+x_{4}^{4}\right) +\frac{1}{2}%
\left( x_{1}^{6}+x_{2}^{6}+x_{3}^{6}+x_{4}^{6}\right) \right]  \nonumber \\
&+&x_{1}x_{2}+x_{1}x_{3}+x_{1}x_{4}+x_{2}x_{3}+x_{2}x_{4}+x_{3}x_{4}\ .
\end{eqnarray}
Table \ref{tab3} compares our eigenvalues with those obtained by Kaluza\cite
{Kaluza} and Chung and Chew\cite{CC07}. Present results are more accurate
than those of Chung and Chew\cite{CC07} for the matrix dimension $%
9^{4}\times 9^{4}$.

\begin{table}[tbp]
\caption{Energy eigenvalues of the system of four coupled sextic anharmonic
oscillators}
\label{tab3}%
\begin{tabular}{cccc}
$N$ & $E_{0}$ & $E_{1}$  & $E_{2}$ \\
\hline
$3^4 \times 3^4$    &  4.133363559 & 6.144782201 & 6.929503230 \\
$5^4 \times 5^4$    &  3.952498514 & 6.276113935 & 7.007385139 \\
$9^4 \times 9^4$    &  3.959409424 & 6.281167988 & 7.016036697 \\
$11^4 \times 11^4$  &  3.959326310 & 6.280902944 & 7.015828402 \\
$13^4 \times 13^4$  &  3.959309441 & 6.280850134 & 7.015787290 \\
$15^4 \times 15^4$  &  3.959305195 & 6.280836518 & 7.015776655 \\
Ref.\cite{CC07}  &  3.960086    & 6.283305    & 7.017863    \\
Ref.\cite{Kaluza}   &  3.959304    &   -         &  -          \\
\end{tabular}
\bigskip \bigskip
\end{table}

\section{CONCLUSIONS}

\label{sec:conclusions}

In this paper we have extended the variational collocation approach
developed earlier in one dimension\cite{Amore07a} to the solution of the
Schr\"{o}dinger equation for systems of coupled oscillators or single
oscillators in more than one dimensions. We have applied our method to
several examples previously considered in the literature and obtained
remarkably accurate results for all of them. In particular, we have shown
how to improve the rate of convergence of our approach by means of nonlinear
variational parameters. We underline some of the virtues of our approach:
its application is straightforward and not limited to polynomial potentials;
the construction of the Hamiltonian matrix does not involve the numerical
calculation of integrals; one obtains its kinetic part for a given grid once
and for all, and store it for applications to problems that differ in the
potential--energy function; the potential part of the Hamiltonian matrix is
diagonal; the Hamiltonian matrix is an \textsl{analytic} function of the
variational parameters which greatly facilitates the numerical determination
of the minimum of its trace.

\appendix

\section{CONSTRUCTION OF THE HAMILTONIAN MATRIX \label{sec:matrix}}

The direct product of $D$ LSF is labelled by $D$ integers $%
(i_{1},i_{2},\ldots ,i_{D})$ where $i_{m}=1,2,\ldots ,M$. In order to
construct a $M^{D}\times M^{D}$ square matrix we have to encode the set $%
(i_{1},i_{2},\ldots ,i_{D})$ into just one integer $\bar{K}_{D}=1,2,\ldots
,M^{D}$. One possibility is to use the following mapping:
\begin{equation}
\bar{K}_{D}=M^{D-1}(i_{1}-1)+M^{D-2}(i_{2}-1)+\ldots +i_{D}  \label{eq:K_D}
\end{equation}

For the purspose of programming, it is useful to have the inverse
relationship that produces the set $(i_{1},i_{2},\ldots ,i_{D})$ for a given
$\bar{K}_{D}$. Notice that we can easily extract $i_{1}$ from $\bar{K}_{D}$
as follows:
\begin{equation}
i_{1}=\left[ \frac{\bar{K}_{D}}{M^{D-1}+\epsilon }\right] +1  \label{eq:i_1}
\end{equation}
where $\left[ x\right] $ stands for the integer part of $x$ and $0<\epsilon
<1$. Proceeding in the same way we have
\begin{equation}
i_{2}=\left[ \frac{\bar{K}_{D}-M^{D-1}(i_{1}-1)}{M^{D-2}+\epsilon }\right] +1
\label{eq:I_2}
\end{equation}
and similar obvious expressions for $i_{3},\ldots ,i_{D-1}$; finally,
\begin{equation}
i_{D}=\bar{K}_{D}-M^{D-1}(i_{1}-1)-M^{D-2}(i_{2}-1)-\ldots -M(i_{D-1}-1)
\label{eq:i_D}
\end{equation}

The LSF label $k$ ranges from $1-N/2$ to $N/2-1$ so that $i=k+N/2$ ranges
from $1$ to $M=N-1$ where $M$ is odd because $N$ is even.

\section{VARIATIONAL METHOD \label{sec:Var_Met}}

In this section we outline some of the variational methods that one commonly
applies to the approximate calculation of the eigenvalues and eigenfunctions
of a given Hamiltonian operator $\hat{H}$:
\begin{equation}
\hat{H}\psi _{n}=E_{n}\psi _{n},\,n=0,1,\ldots  \label{eq:Schrodinger}
\end{equation}

One can insert a variational parameter $\alpha $ (or a set of such
parameters) into a trial function $\varphi $ by means of a unitary operator $%
\hat{T}(\alpha )$: $\varphi (\alpha )=\hat{T}\varphi $. Since $\hat{T}$ is
unitary then $\hat{A}=(\partial \hat{T}/\partial \alpha )\hat{T}^{\dagger }$
is antihermitian: $\hat{A}^{\dagger }=-\hat{A}$. If $\varphi $ is normalized
then $\hat{T}\varphi $ will also be normalized and the variational method
leads to the hypervirial theorem for the optimal variational function\cite
{FC87}:
\begin{equation}
\frac{\partial }{\partial \alpha }\left\langle \varphi (\alpha )\right| \hat{%
H}\left| \varphi (\alpha )\right\rangle =\left\langle \varphi (\alpha
)\right| [\hat{H},\hat{A}]\left| \varphi (\alpha )\right\rangle =0
\label{eq:HT}
\end{equation}

If we have an orthonormal set of functions $\{\phi _{n}\}$ we obtain a
variational set exactly as before: $\{\phi _{n}(\alpha )=\hat{T}\phi _{n}\}$
and then apply the Rayleith--Ritz variational method that leads to
\begin{equation}
\sum_{j=0}^{N-1}\left[ H_{ij}(\alpha )-W\delta _{ij}\right] c_{j}=0
\label{eq:secular}
\end{equation}
The approximate energies $W_{k}(\alpha )$, $k=0,1,\ldots ,N-1$ depend on the
variational parameter $\alpha $. Since $W_{k}(\alpha )>E_{k}$ then it seems
reasonable to obtain $\alpha =\alpha _{k}^{opt}$ such that
\begin{equation}
\left. \frac{\partial W_{k}}{\partial \alpha }\right| _{\alpha =\alpha
_{k}^{opt}}=0  \label{eq:Var_Wk}
\end{equation}
However, this procedure commonly requires considerable coputer time. For
this reason many authors resort to
\begin{equation}
\left. \frac{\partial H_{00}}{\partial \alpha }\right| _{\alpha =\alpha
_{0}}=0  \label{eq:Var_H00}
\end{equation}
that clearly reduces computer time considerably but is mainly suitable for
the ground state.

An alternative approach is to minimize the trace of the Hamiltonian matrix
\begin{equation}
\frac{\partial }{\partial \alpha }\sum_{j=0}^{N-1}H_{jj}(\alpha )=\frac{%
\partial }{\partial \alpha }\sum_{j=0}^{N-1}W_{j}(\alpha )=0
\label{eq:Var_trace}
\end{equation}
that provides a balanced approximation to all the states considered in the
Rayleigh--Ritz calculation.

It is worth noticing that we can insert the adjustable parameters either
into the trial function $\varphi $ or into the Hamiltonian operator
according to $\left\langle \hat{T}\varphi \right| \hat{H}\left| \hat{T}%
\varphi \right\rangle =\left\langle \varphi \right| \hat{T}^{\dagger }\hat{H}%
\hat{T}\left| \varphi \right\rangle $.

Consider a general dimensionless Hamiltonian operator of the form $\hat{H}(%
\mathbf{x},\mathbf{p})$ where $\mathbf{x}$ is a vector with components $\hat{%
x}_{i}$ and $\mathbf{p}$ is a vector with the conjugate momenta $\hat{p}_{j}$
such that $[\hat{x}_{j},\hat{p}_{k}]=i\delta _{jk}$, where $i,j=1,2,\ldots
,D $. In what follows we consider a variational basis set of the form\cite
{FC87}
\begin{equation}
\varphi _{n}(\mathbf{x})=\hat{T}\phi _{n}(\mathbf{x})=\sqrt{\det (\mathbf{C})%
}\phi _{n}(\mathbf{Cx})  \label{eq:var_basis}
\end{equation}
where $\mathbf{C}$ is a conveniently chosen square matrix that provides a
set of variational parameters. Suppose that $\alpha $ is one of the
adjustable parameters in the matrix $\mathbf{C}$; then, it is not difficult
to prove that the antihermitian operator $\hat{A}$ is given by
\begin{equation}
\hat{A}=\frac{1}{2}\frac{\partial \ln \det \mathbf{C}}{\partial \alpha }%
+\sum_{i}\sum_{j}\left( \mathbf{C}^{-1}\frac{\mathbf{C}}{\partial \alpha }%
\right) _{nm}x_{m}\frac{\partial }{\partial x_{n}}  \label{eq:Aop}
\end{equation}
It is worth noticing that there is one antihermitian operator $\hat{A}$ for
each variational parameter $\alpha $ although this equation does not reflect
this fact explicitly.

As a first simple example consider the scaling transformation given by $%
C_{ij}=\alpha _{i}\delta _{ij}$ that leads to the scaling operators
\begin{equation}
\hat{A}_{i}=\frac{1}{2\alpha _{i}}+\frac{1}{\alpha _{i}}x_{i}\frac{\partial
}{\partial x_{i}},\,i=1,2,\ldots ,D  \label{eq:Ai_scaling}
\end{equation}
If we rewrite these equations in terms of momenta in the coordinate
representation $\hat{p}_{j}=-i\partial /\partial x_{j}$ and rewrite the
scaling transformation $x_{j}^{\prime }=\alpha _{j}x_{j}$, $\hat{p}%
_{j}^{\prime }=\alpha _{j}^{-1}\hat{p}_{j}$ in terms of creation and
anihilation operators $\hat{a}_{j}=(\hat{x}_{j}+i\hat{p}_{j})/\sqrt{2}$, $%
\hat{a}_{j}^{\dagger }=(\hat{x}_{j}-i\hat{p}_{j})/\sqrt{2}$, $\hat{b}_{j}=(%
\hat{x}_{j}^{\prime }+i\hat{p}_{j}^{\prime })/\sqrt{2}$, $\hat{b}%
_{j}^{\dagger }=(\hat{x}_{j}^{\prime }-i\hat{p}_{j}^{\prime })/\sqrt{2}$
then we conclude that the Bogoliubov transformation $\hat{b}_{j}=\left( \hat{%
a}_{j}-t_{j}a_{j}^{\dagger }\right) /\sqrt{1-t_{j}^{2}}$, $\hat{b}%
_{j}^{\dagger }=\left( \hat{a}_{j}^{\dagger }-t_{j}a_{j}\right) /\sqrt{%
1-t_{j}^{2}}$ is equivalent to the well--known scaling discussed above. We
clearly appreciate that the fashionable two--step approach\cite{HC84,CC07}
is simply the scaled Rayleigh--Ritz variational method that has proved
useful since long ago in molecular calculations\cite{L59}.

If the coordinate transformation is orthogonal $\mathbf{C}^{T}=\mathbf{C}%
^{-1}$ then the matrix $\mathbf{A=C}^{T}\partial \mathbf{C}/\partial \alpha $
is antisymmetric $\mathbf{A}^{T}=-\mathbf{A}$, and
\begin{equation}
\hat{A}=\sum_{m}\sum_{n>m}A_{mn}\left( x_{m}\frac{\partial }{\partial x_{n}}%
-x_{n}\frac{\partial }{\partial x_{m}}\right)  \label{eq:A_op_rot}
\end{equation}
If the problem is described by $D$ independent coordinates $x_{i}$ then we
have $D(D-1)/2$ components of the angular--momemtum operator like the one
between parenthesis in Eq.~(\ref{eq:A_op_rot}).

As an example we consider the Hamiltonian operator studied by Witwit\cite
{Witwit93}:
\begin{eqnarray}
\hat{H} &=&\frac{1}{2}\left( p_{x}^{2}+p_{y}^{2}+p_{z}^{2}\right) +\frac{1}{2%
}\left( x^{2}+y^{2}+z^{2}\right)  \nonumber \\
&&+\lambda \left( a_{xx}x^{4}+a_{yy}y^{4}+a_{zz}z^{4}\right.  \nonumber \\
&+&\left. 2a_{xy}x^{2}y^{2}+2a_{xz}x^{2}z^{2}+2a_{yz}y^{2}z^{2}\right)
\label{eq:H_Wit_3D}
\end{eqnarray}
In this case we may try three rotation parameters that will lead to the
hypervirial relations\cite{FC87}
\begin{equation}
\left\langle \varphi \right| [\hat{H},\hat{L}_{u}]\left| \varphi
\right\rangle =0,\,u=x,y,z  \label{eq:HT_rot}
\end{equation}
If $\varphi $ is a product of harmonic--oscillator eigenfunctions the
hypervirial theorems are satisfied by the unrotated functions. In other
words, coordinate rotation does not appear to provide useful variational
parameters because of the symmetry of this problem. This is obviously true
if we use the prescriptions (\ref{eq:Var_Wk}), (\ref{eq:Var_H00}) and (\ref
{eq:Var_trace}). However, as shown in Sec~\ref{sec:D=3}, the rotation
transformation is useful to improve present LSF calculations.

\verb''\section*{References}
\verb''

\end{document}